# Operation of the 56 MHz Superconducting RF Cavity in RHIC with Higher Order Mode Damper

Q. Wu[1], S. Belomestnykh[2,3], I. Ben-Zvi[1,3], M. Blaskiewicz[1], T. Hayes[1], K. Mernick[1], F. Severino[1], K. Smith[1], A. Zaltsman[1]

[1]BNL, Upton, NY 11973, USA
[2]Fermilab, PO Box 500, Batavia, IL 60510, USA
[3]Stony Brook University, Stony Brook, NY, 11790, USA

*Abstract*

A 56 MHz superconducting RF cavity was designed and installed in the Relativistic Heavy Ion Collider (RHIC). It is the first superconducting quarter wave resonator (QWR) operating in a high-energy storage ring. We discuss herein the cavity operation with Au+Au collisions, and with asymmetrical Au+He3 collisions. The cavity is a storage cavity, meaning that it becomes active only at the energy of experiment, after the acceleration cycle is completed. With the cavity at 300 kV, an improvement in luminosity was detected from direct measurements, and the bunch length has been reduced. The uniqueness of the QWR demands an innovative design of the higher order mode dampers with high-pass filters, and a distinctive fundamental mode damper that enables the cavity to be bypassed during the acceleration stage.

## I. Introduction

RHIC was designed with accelerating and storage RF systems to operate at harmonic number $h = 360$, and $h = 7 \times 360 = 2520$, which correspond to frequencies of 28.15 MHz and 197.05 MHz respectively [15]. The five 197 MHz copper cavities installed in each ring are used to store bunches for up to 10 hours after they have been accelerated to the top energy. With the current RHIC operational settings, the typical full length of the 100 GeV Au beam is 9.7 ns with the 28 MHz acceleration RF system. The bucket length of the relatively high frequency storage cavities is 5.1 ns, which is not sufficient to accommodate the ion bunches without going through the RF gymnastics of bunch rotation, or re-bucketing. However, this re-bucketing process causes an increase in longitudinal emittance due to nonlinearity and hardware complications. The longitudinal emittance increase leads to 30% of the particles being spilled into neighboring buckets, which hardly contribute to the luminosity at collision.

A storage cavity at a frequency of 56.3 MHz was proposed in 2007 to provide adequate RF acceptance to long bunches without re-bucketing [2]. The longitudinal acceptance using the 56 MHz cavity at RHIC is 6.4 eV·s/u for 100 GeV/u gold ions and 15.5 eV·s/u for 250 GeV protons, in both cases much larger than the beam emittance [3]. With the same longitudinal emittance, the full bunch length with the 56 MHz RF system may be suppressed to below the bucket size of the 197 MHz cavity [3]. As was shown from direct luminosity measurements during RHIC operation in 2014, the combination of the 56 MHz and 197 MHz RF systems raised the luminosity within the vertex of the detectors, which will be discussed in the later section of this paper.

The 56 MHz RF cavity installed in RHIC is the first quarter wave superconducting resonator for storage rings, which supports a high gap voltage to above 2 MV. With such an advantage, the number of cavities can be reduced to only one, which is further shared by both rings, and therefore has somewhat lowered the RHIC impedance. The cavity operates with the beam traversing through its axis. This design provides a compact size making feasible its installation in the existing tunnel.

The 56 MHz cavity is beam driven, although a maximum of 1 kW RF power is provided from an amplifier for controlling its amplitude. After the beam has been accelerated to top energy, the cavity is tuned towards the beam frequency from below. Since the cavity operates above the transition energy, Robinson instability [4, 5] is avoided. Nevertheless, external RF power is necessary to ensure stability of its amplitude against ponderomotive instabilities [6, 7, 8].

The instability of the coupled bunch of the ring-cavity system was studied [9], and damping is provided accordingly to the fundamental and higher order modes to the specified thresholds for passing through the transition energy and at store.

## A. Cavity

The 56 MHz SRF cavity is a quarter-wave resonator with beam passing along its axis of symmetry, as shown in Figure 1. The cavity is located in the common section of the RHIC, and shared by ion bunches from both rings. Serving both rings with one cavity provides costs benefits for construction and operation [10]. To achieve identical longitudinal dynamic effects for both beams, the difference in RF phase must be equal to $\pi$. The two colliding beams are synchronized at the interaction point (IP). Thus the cavity should be installed such that the accelerating gap is located at $(n + 1/4)\lambda$ away from the IP. The wavelength at a frequency of 56.3 MHz is $\lambda = 5.33$ m. The available position in RHIC is located at $1.25\lambda$, which is 6.66 m from the IP. The parameters of the cavity operating at 2 MV are shown in Table 1.

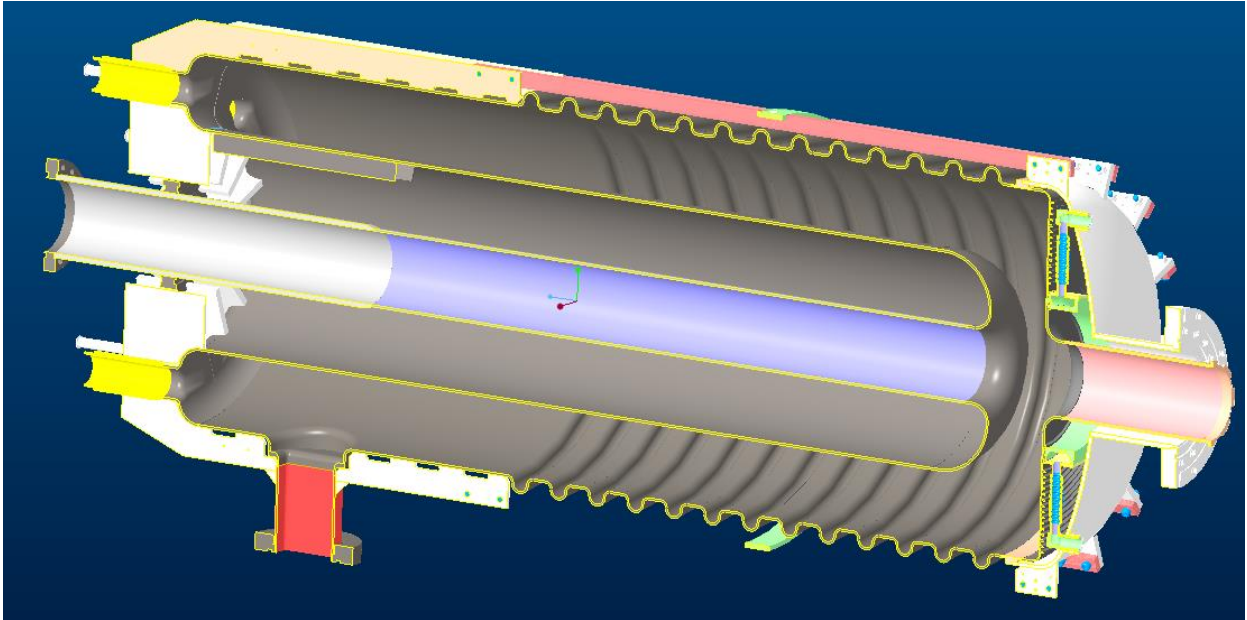

Figure 1: 56 MHz SRF cavity with a total length of 1.3 meters.

Table 1: Cavity parameters

| Parameter | Unit | Value |
|---|---|---|
| Frequency | MHz | 56.30 |
| Length | cm | 134.88 |
| Gap | cm | 8.45 |
| Outer diameter | cm | 50.5 |
| Aperture | cm | 10.0 |
| Operating temp | K | 4.5 |
| R/Q | Ω | 80.5 |
| $Q_0$ | | $1.95 \times 10^9$ |
| $E_{acc}$ | MV/m | 23.5 |
| $E_{peak}/E_{acc}$ | | 1.77 |
| $B_{peak}/E_{acc}$ | mT/MV/m | 3.81 |

The cavity is in line with a standard quarter wave resonator. Considering of the height limit inside the RHIC tunnel to accommodate the full cryomodule and sufficient headroom for its cryogenic system, the cavity diameter was set to be less than 50 cm. By optimizing the shape locally near the gap, the cavity design aimed on both decreasing of the ratio of the peak electric field to the accelerating field and increasing the sensitivity of the tuning plate [11].

*Tuning*

The cavity is tuned by pushing and pulling a flat tuning plate at the accelerating gap, as shown in Figure 1. To change the frequency, the flat surface of the cavity is connected to the cylindrical body with an intentional 1 mm thin Nb membrane, as shown in Figure 2. This vulnerable thin connecting section is separated from other environments by a torispherical dome, so that the tuning plate is not exposed to the helium or vacuum pressure. The pressure is always kept balanced on both side of the tuning plate to protect the thin section.

Before the final chemical treatment, a pre-tuning step was processed by plastically change the tuning plate connecting section. The pre-tuning process pulled the tuning plate with the cylindrical cavity body fixed and increased the length of the cavity by 2.8 mm, which corresponds to a frequency increase of 72.5 kHz. During the operation phase, the entire elastic tuning range is 46.25 kHz, which corresponds to 3 mm motion of the tuning plate. During operation, the cavity is tuned 34 kHz below the beam resonance in order to minimize the beam excited cavity voltage to a negligible level. During operation, a piezo tuner is employed to provide a fine tune and a fast response in addition to the stepper tuner, and potentially compensate detuning due to microphonic noise.

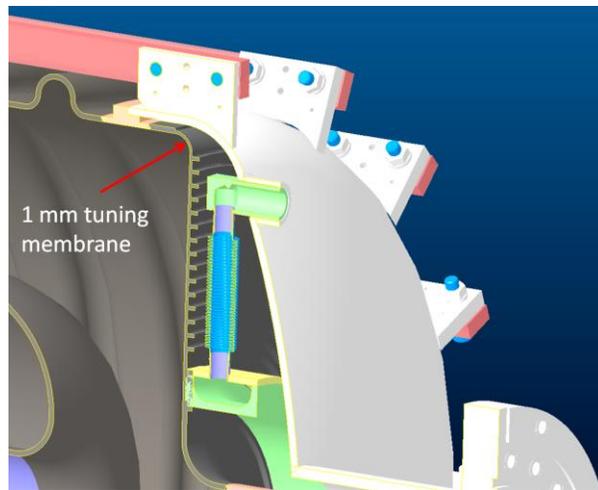

Figure 2: Tuning section of 56 MHz cavity with push-pull on 1mm Nb membrane for changing the cavity volume.

*Multipacting Solution*

As the cavity has traditional coaxial shape, multipacting is easily excited between the parallel walls of inner and outer conductors. For suppressing multipacting, thirteen corrugations were implemented on the outer wall [12]. When entering the corrugation region, the path of the secondary electrons deviates from their resonating condition, and therefore, there is no local accumulation of large quantities of electrons. The corrugation is 2 cm in diameter, with 6 cm spacing. Simulations showed that only the first 70 cm from cavity gap has an electric field strong enough for the resonance condition. Before operating the cavity, we encountered multipacting at low field corresponding to 1 kV cavity voltage. This two point multipacting event happens in the corner of the head plate and the nearby cavity cylinder, and it is conditioned away with 1 kW amplifier and external Q of the fundamental mode coupler of 1e6. 90 to 140 kV gap voltage. With the 1 kW amplifier running in the CW mode, the multipacting zone was conditioned away after 30 minutes. Following its conditioning, the cavity never encountered multipacting during its operation with beam.

## B. Couplers

The cavity does not have a sufficient tuning range to follow a large change in frequency during the RHIC energy ramp, so it is tuned up to the resonance frequency only after reaching the store energy. In addition to fully detuning the cavity with the tuning plate, a fundamental mode damper (FMD) is inserted into the cavity from a rectangular port opened on the side. The fully inserted FMD is shown in Figure 3.

The FMD location is chosen with its center axis at 114.3 cm from the tuning plate. The strong coupling of this coupler interferes with the cavity frequency along the insertion and extraction process, therefore the design has carefully chosen the location to have a minimum frequency shift by balancing the capacitive and inductive effects. The loaded Q and the frequency change due to the motion of the FMD is shown in . A combination of the tuning plate and fully inserted FMD detunes the cavity frequency by 68.5 kHz from the beam resonance. When FMD is fully inserted, the cavity loaded Q is 309. The cavity voltage excited by the full intensity RHIC beam after the large detune and heavy damping, is less than 6 kV.

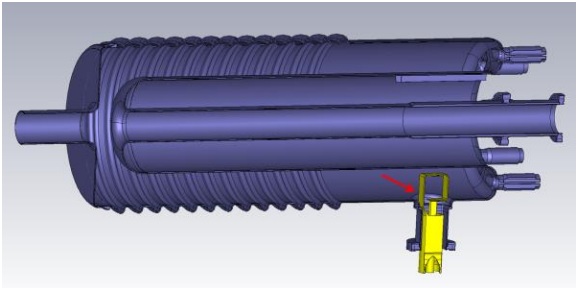

Figure 3: The fundamental mode damper fully inserted into the cavity.

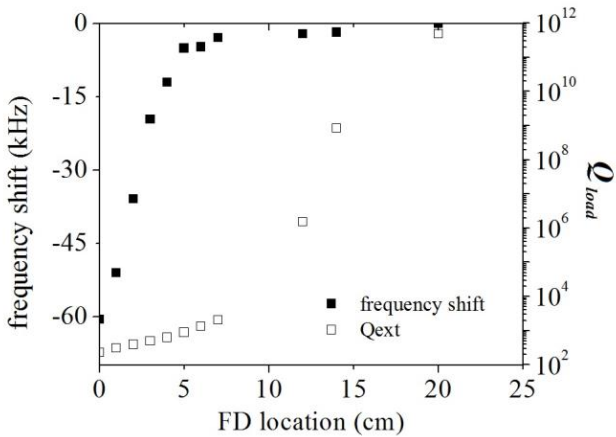

The 56 MHz cavity is a passive resonator, whose energy is built up by beam loading and supplied from the 28 MHz main accelerating system. In addition, an external amplifier with a maximum power of 1 kW is connected to the cavity via a fundamental power coupler (FPC). The loop-shaped FPC, located in the maximum magnetic field region, is shown in Figure 4. Insertion of the FPC is variable via a stepper motor, and the external $Q$ changes by two orders of magnitude. Due to small losses of the superconducting cavity, the amplifier power alone is sufficient for driving the cavity to above 2 MV on resonance and initial conditioning. During conditioning, the FPC is parked in its fully inserted position with an external $Q$ of $2.2\times10^6$ to maximize the transmitted RF power to the cavity.

During operation, the FPC couples RF power into the cavity to compensate the amplitude change in the gap voltage, which then is used as the real time correction of the voltage. Less than 10 watts from the amplifier is required to maintain the stability of the amplitude and the phase.

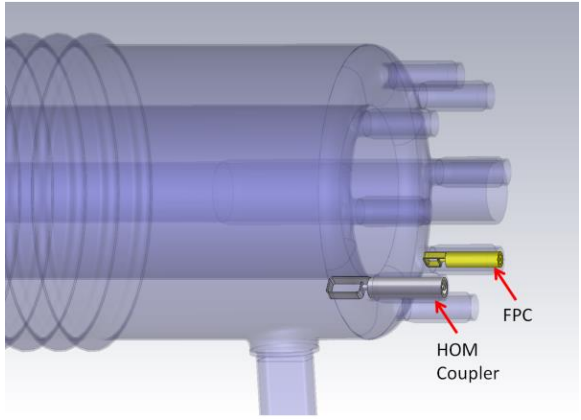

Figure 4: FPC (yellow) and HOM coupler (green) locations in the cavity.

Higher order modes (HOMs) of the cavity are damped through an HOM loop-shaped coupler, as shown in Figure 4. The FMD port breaks the symmetry of the cavity, therefore location of the HOM coupler is specifically chosen for attaining a balanced optimum damping of HOMs. For all HOMs below 700 MHz, we measured that the coupler damps power of all dipole modes by 5 orders of magnitude and of quadrupoles by 6 orders of magnitude.

The HOM coupler loop couples to all modes, including the fundamental mode. To avoid dangerously large amount of power from the fundamental mode being sent into the HOM coupler load, a Chebyshev high-pass filter was connected to the coupler loop. Frequency of the first HOM is 168 MHz. Thus the frequency difference between the fundamental mode and the first HOM is greater than 100 MHz, which simplified the filter design. The HOM coupler and filter are discussed in the next section with details.

## C. HOM couplers with high pass filter

As of the HOM analysis for the 56MHz SRF cavity, we define the R/Qs as the following:

$$\frac{R}{Q}_{monopole} = \frac{V^2}{\omega U} = \frac{\left|\int E_{z,axis} e^{ikz} dz\right|^2}{\omega U}$$

$$\frac{R}{Q}_{dipole} = \frac{1}{k^2 d^2}\frac{V_d^2}{\omega U} = \frac{1}{k^2 d^2}\frac{\left|\int E_{z,d} e^{ikz} dz\right|^2}{\omega U}$$

$$\frac{R}{Q}_{quadrupole,sextupole} = \frac{V_d^2}{\omega U} = \frac{\left|\int E_{z,d} e^{ikz} dz\right|^2}{\omega U}$$

where $\omega$ and $U$ are the frequency and stored energy of the specified mode, $k = \omega/c$ is the wave number, $z$ is the direction along the cavity symmetric axis, $E_{z,axis}$ and $E_{z,d}$ are the electric fields in $z$ direction on the axis and with a offset of $d$. For the 56MHz cavity, we chose $d = 1cm$.

The $R/Q$s of the monopoles and dipoles are, as plotted in Figure 5, distributed between 0 to 100, while the $R/Q$s of the quadrupoles and sextupoles are all below $10^{-2}$. Therefore, while the effective damping for all the modes is important, the monopole and dipole modes are more critical compared to the other two types.

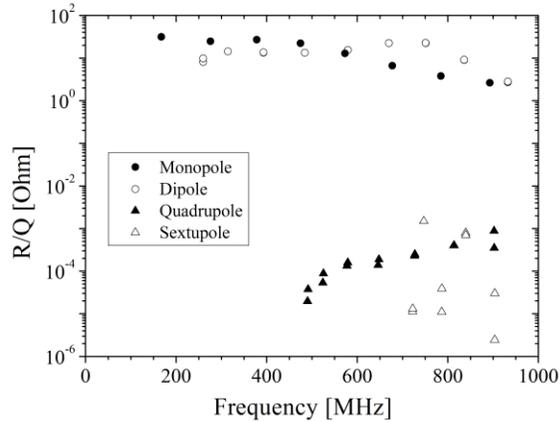

Figure 5: R/Q of all cavity modes.

## HOM Coupler Loop

We chose the chemical cleaning ports as the insert ports of our HOM couplers to prevent extra penetrations to the cavity. All of the chemical cleaning ports are located at the peak magnetic field region of the cavity, thus the HOM coupler design adopted a rectangular loop with an opening area of $6cm \times 2.88cm$, Figure 6. The width of the loop was set as $2cm$ and its thickness at $0.3cm$. The size of the coupler loop was designed to provide a sufficient inductive coupling, while also allowing enough clearance through the 4 cm diameter opening of the port. To limit the peak surfacefield of the HOM coupler to below 80 mT during operation, the thickness of the loop is set to be below 0.5 cm. At the meantime, the width of the loop is chosen for the loop to contain enough material for thermal conduction. To limit the peak surface field of the HOM coupler to below 80 mT during operation, the thickness of the loop is set to be below 0.5 cm. At the meantime, the width of the loop is chosen for the loop to contain enough material for thermal conduction.

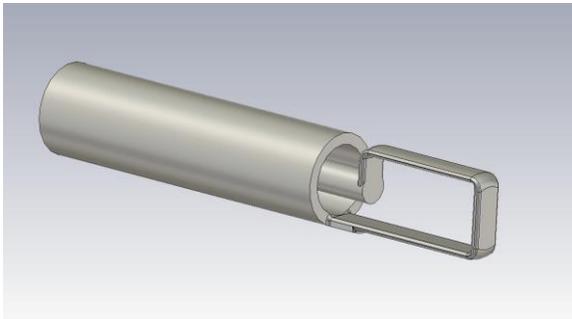

Figure 6: HOM coupler loop.

Since the size of the damper loop is limited by the port opening, the initial design of the cavity was with multiple HOM dampers installed. Out of the 8 ports, 4 of them are occupied by the fundamental power coupler, pick-up probe, and two IR sensors for multipacting detection. Thus we are limited to 4 HOM dampers with the size described previously to fulfil the damping requirement of the 56MHz SRF cavity.

All 8 chemical ports are distributed uniformly around the cavity end. The choice of the ports selected for HOM coupler insertion is determined by the damping effect of each scheme. As shown in Figure 7, 4 couplers are inserted 90 degrees apart from each other in a symmetrical pattern. All the loops are oriented radially as shown in the figure to maximize the effective coupling area.

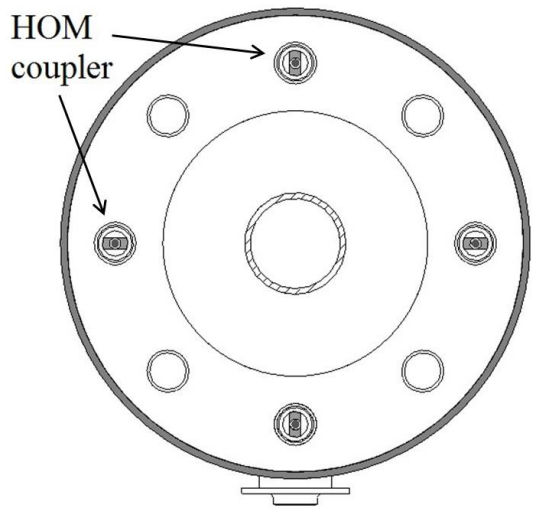

Figure 7: End view of cavity with symmetrical distribution of the HOM couplers.

On the cavity, the FD port breaks the 2D symmetry along the beam axis, therefore the orientation of the multipole modes are determined accordingly. As for the dipole modes, the configurations are either aligned with or perpendicular to the FD port. As shown by the dots in Figure 8, the 4 HOM couplers can effectively damp all the dipole modes. The triangles representing the quadrupole modes in Figure 5 are separated into two sets of damping results, external Q ($Q_{ext}$) value of below 105 and above 108. The low $Q_{ext}$ values are associated with the quadrupole modes that are oriented in the same configuration as of the couplers, while high $Q_{ext}$s are for the orthogonal modes. High $Q_{ext}$ indicate the damping of the corresponding modes is not sufficient.

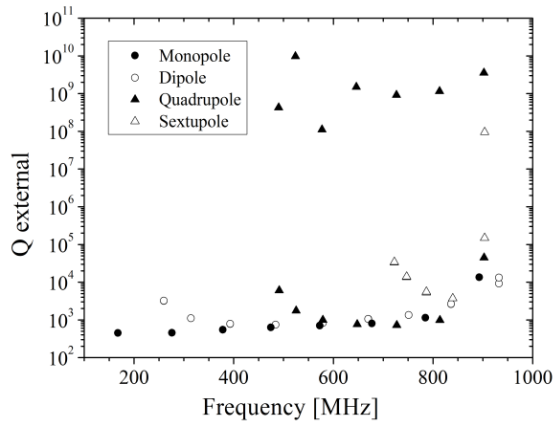

Figure 8: HOM coupling from symmetrically distributed couplers.

To provide coupling to orthogonal quadrupole HOMs, the four dampers must be placed asymmetrically, as shown in Figure 6. In this scheme, the bottom HOM coupler is rotated clockwise by 45 degrees, so that it will have strong coupling with the quadrupole modes missed by the other three couplers. Figure 7 shows that all the $Q_{ext}$ of the coupler ports at such assymetrical configuration of the HOM couplers are under 105. As a compensation of the quadrupole coupling improvement, the coupling to the vertical dipole modes will decrease, but the difference is small.

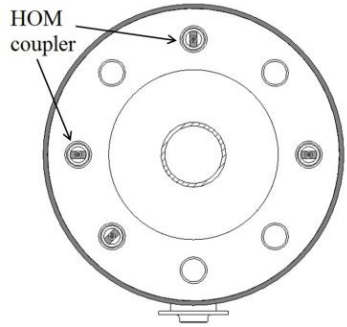

Figure 9: End view of cavity with asymmetrical distribution of the HOM couplers.

As shown in Figure 10, sufficient damping is provided to all the modes, while damping to the monopole and dipole modes is more than an order of magnitude stronger than the other two type of modes.

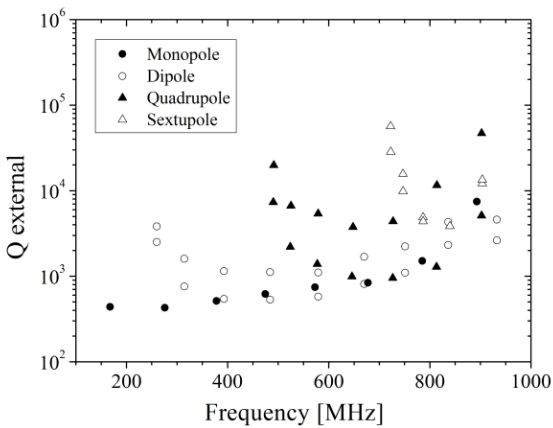

Figure 10: HOM coupling from asymmetrically distributed couplers.

All HOM coupling is analysed with MicroWave Studio 3D simulation [2].

## *High-Pass Filter*

The HOM dampers will remain in the cavity during operation; therefore all modes, including the fundamental mode, will be coupled out. According to our MWS simulation, the loaded Q of the fundamental mode of the cavity will drop to about $10^3$. Compared to the designed loaded Q of $4 \times 10^7$, this will send a dangerously large amount of power into the HOM damper load. The frequency of the first HOM is 168MHz, and thus, the frequency difference between the fundamental mode and the first HOM is greater than 110MHz. To address such a large separation, we designed and incorporated a high-pass filter into the circuit to reflect the fundamental mode.

In addition to reflecting the fundamental mode, the filter must contain a vacuum seal to separate the ultra high vacuum in the cavity from the insulation vacuum. Because the cavity is superconducting, the design of the filter must be such to also prevent excessive heat transfer into the HOM coupler and so raising the temperature in the vicinity of the cavity. Furthermore, since the ports for the HOM couplers are needed to allow access for chemically treating the cavity, the entire HOM coupler assembly must be detachable from the cavity.

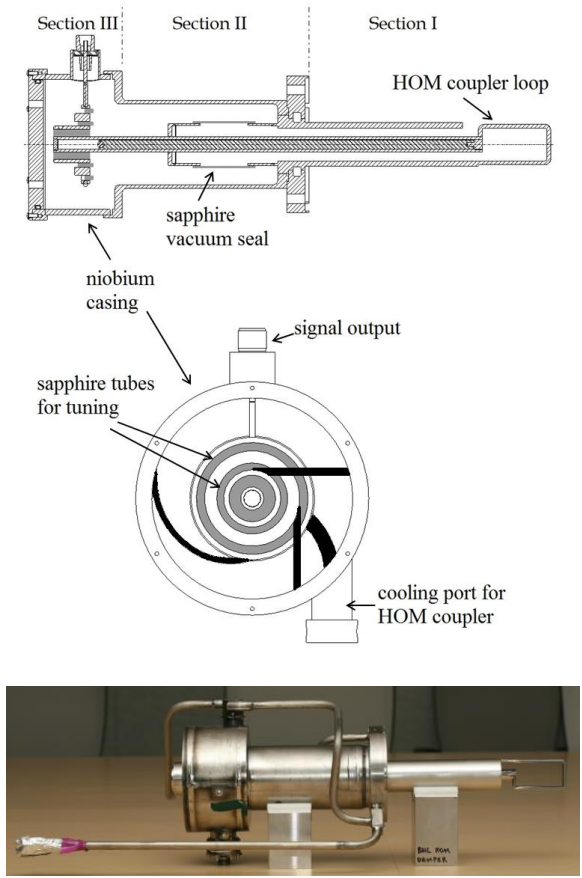

Figure 11: Top: Cross section view of the HOM coupler with high-pass filter assembly. Middle: Side view of the high pass filter. Bottom: HOM Coupler with high-pass filter and cooling channels fabricated at Thomas Jefferson Lab.

We designed the filter as a three-stage high-pass Chebyshev T-Type filter to obtain a steep roll-off and minimize the error between the designed and the actual object, Figure 11. In the cross sectional view, the assembly has been separated into three sections. Section I is the HOM loop that penetrates completely into the cavity, along with the coaxial structure for lossless transport of the RF power. Section II is the seal for the cavity's ultra-high vacuum. The connection between center and outer conductors of the coaxial line is joined by a sapphire tube brazed onto two niobium cuffs, which then are welded to the two conductors. The sapphire tube is 6.35 cm long, and can withhold more than 3 MV. Section III is the high pass filter structure. Three coaxial sapphire rings, shown in grey, with niobium spacers constitute the three capacitors in the design; the four inductors are niobium rods, shown in black, with carefully chosen lengths and locations. All the niobium rods start on one of the niobium cuffs, with their other end being welded on the outer can of the high pass filter. A side view middle of Figure 11 shows all the sapphire rings and niobium rods for the filter. We selected sapphire due to its low loss tangent, to reduce heat dissipation in the HOM damper caused by dielectric heating. To assure better conductive cooling of the loop, we designed the filter with an inductor first, as shown in the lumped circuit diagram and the assembly after fabrication in Figure 12. The center conductor of the coaxial structure and the first inductor both are fabricated with niobium tubes, and high RRR copper rods are inserted into these tubes leaving a clearance for thermal expansion. The copper rods are welded together at the joint to serve as a cooling path for the HOM coupler loop.

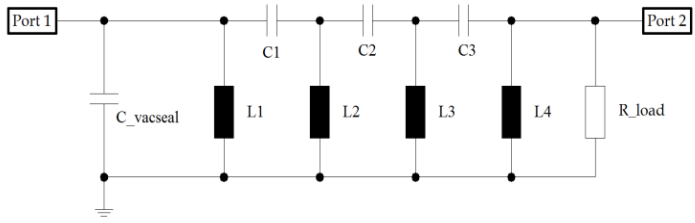

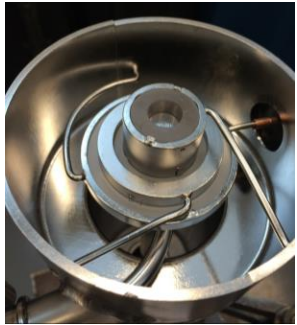

Figure 12: Top: Simplified RLC circuit for the high-pass filter. Bottom: High-pass filter section in the HOM coupler.

We calculated the values of the lumped RLC circuit individually according to the actual components in the design; Table 2 gives the results. Here, we also took into account the different dielectric constant for sapphire in its radial and azimuthal direction of the anisotropy axis at 4.2 K. Port 1 corresponds to the input of coaxial structure after the coupler loop, while port 2 corresponds to the signal output port.

Table 2: Components of the equivalent circuit for the high pass filter.

| Component | Unit | Capacitance/Inductance |
|---|---|---|
| $C_{vacseal}$ | pF | 0.41 |
| $C_1$ | pF | 23.1 |
| $C_2$ | pF | 24.9 |
| $C_3$ | pF | 13.3 |
| $L_1$ | nH | 40.3 |
| $L_2$ | nH | 40.1 |
| $L_3$ | nH | 37.5 |
| $L_4$ | nH | 75.8 |

In Figure 13, we depict the response of the circuit from DC to 1GHz with a solid line. The steep rise ensures the fundamental mode reflection at 56 MHz with effective damping of the first HOM at 168 MHz. The simulated transport parameter of the 3D high pass filter model for all the cavity modes below 1 GHz is also shown on the same figure in a solid square. Initially, the simulation of the model and the lumped circuit match well, but differences develop in modes around 830 MHz. They are due to vacuum gaps in the assembly that add in extra capacitance components in the high frequency region. However, both results show more than a 70 dB reduction in the fundamental mode power and less than 10 dB reduction for all HOMs except the four modes around 830 MHz; nevertheless, they display a reduction of less than 25 dB.

During operation, only one HOM coupler assembly has been fabricated and leak checked due to limited time window. This HOM coupler was installed on the cavity at the 8 o'clock location in Figure 9.

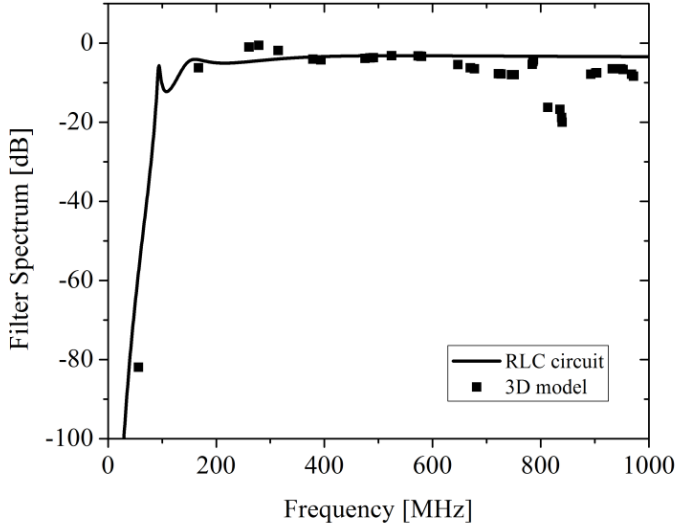

Figure 13: S21 spectrum of the 56 MHz cavity HOM high-pass filter.

## II. RHIC Operation

### A. Without HOM Coupler

The cavity was tested with no HOM coupler installed. The cavity operated with 11 bunches of proton at a total intensity of $25 \times 10^9$. The proton bunches were injected in an evenly distributed 12-bunch pattern, which provided maximum RF field damping between bunches. After extracting the fundamental mode damper, the cavity was slowly tuned towards the beam resonance from 40 kHz below the target frequency.

The tuning ranges of the HOMs from the tuner are much larger than the fundamental mode. All monopole modes below 500 MHz, as listed in Table 3, were excited across the full frequency sweep. Significant beam instability was observed from the effect of these HOMs as they are excited over the tuning process.

Table 3: Excited monopoles measured during cavity tuning towards beam resonance.

| Frequency [MHz] | R/Q [Ohm] | Full Tuning Range [kHz] |
|---|---|---|
| 165.9 | 34 | 848 |
| 274.0 | 29 | 299 |
| 377.1 | 30 | 447 |
| 474.8 | 25 | 420 |

The cavity frequency was tuned and parked at less than 1 kHz away from the beam resonance line. The final bunch length of the proton beam increased by 2.7 folds of the original size, and significant beam loss occurred at each HOM excited frequency, as shown in Figure 14. The beam profile decay was recorded through wall current monitor after passing all HOM lines, as shown in Figure 15.

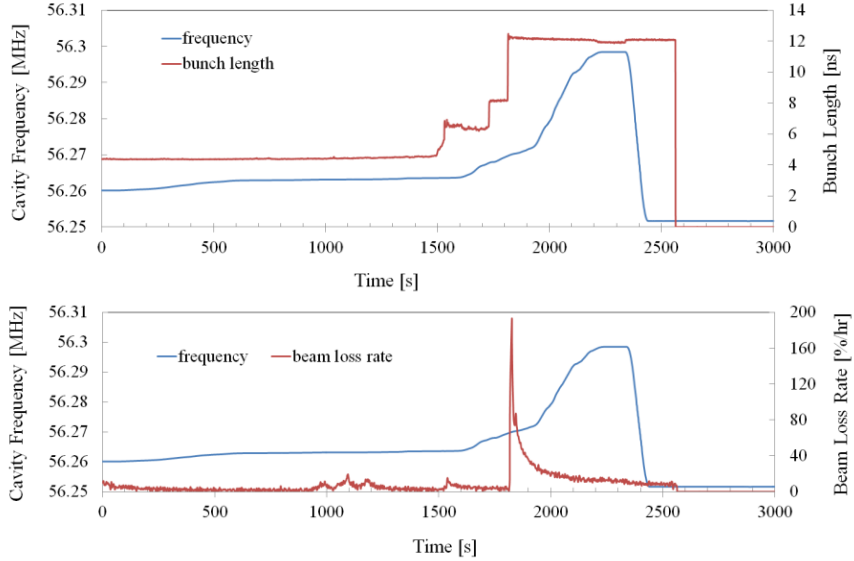

Figure 14: Both the bunch length (top) and beam loss rate (bottom) increase as cavity frequency swept through full tuning range.

No dipole modes or above were excited during the cavity operation, and the HOM power beyond 500 MHz was negligible. The RF amplifier was turned off during the cavity operation.

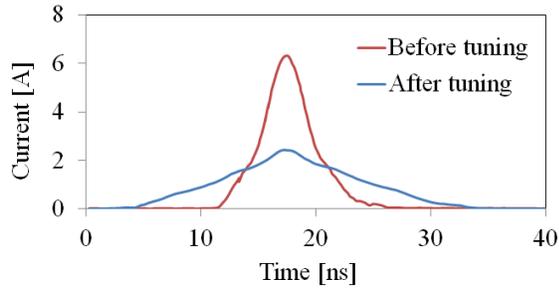

Figure 15: Beam profile decay from 40 kHz cavity tuning towards the beam resonance line.

## B. With HOM Coupler

The cavity was commissioned with one HOM coupler installed. Viewing from the coupler port side, the HOM coupler is 45 degrees clock wise from the fundamental mode damper. The cavity gap voltage was limited to 300 kV, due to the HOM coupler thermal quench, compared with the planned operational voltage of 2 MV. Table 4 lists the typical beam parameters.

Table 4: RHIC operation parameters

|  | Au+Au | He3+Au | |
|---|---|---|---|
|  | Au(b/y) | He(b) | Au(y) |
| Intensity ($\times 10^9$) | 180 | 450 | 180 |
| Energy (GeV) | 100 | 103 | 100 |
| Bunch length w/o 56MHz Cavity (ns) | 4.5 | 6.0 | 4.5 |

Signal was measured from the HOM coupler during operation of the cavity, with full RHIC beam and cavity voltage set at 300 kV. From the results shown in Table 5, the RF performance was very close to the design after fabrication, tuning, and cool down to 4K.

Table 5: Power comparison of each mode from the HOM coupler output during operation.

| Mode # | Frequency [MHz] | Qext_sim | Qext_measured | P_sim [W] | P_measured [W] |
|---|---|---|---|---|---|
| 1 | 56.3126 | 4.1e10 | 3.46E+10 | 1.0 | 2.523481 |
| 2 | 166.233 | 2031 | 6898 | 3.59 | 4.064433 |
| 3 | 273.7679 | 2593 | 6313 | 0.023 | 0.033113 |
| 4 | 376.7558 | 3359 | 17349 | 0.0016 | 0.000871 |
| 5 | 474.2038 | 4142 | 2131.5 | 0.00634 | 0.000955 |
| 6 | 573.541 | 6895 | 11490 | 0.000198 | 0.000891 |
| 7 | 678.9392 | 4850 | 13497 | 0.000063 | <<< |

Even with the six-fold lower voltage in the cavity during the Au + Au run, a reduction in bunch length and an increase in luminosity, reflected by the Zero Degree Calorimeter (ZDC) rate of the detectors, are observed clearly as the cavity is turned on, as illustrated in Figure 16. The bunch length of both beams shrinks by about 4.5%, when the 56 MHz cavity is turned on. The hourglass effect, which refers the increase of the betatron function and the size of the transverse beam away from the IP therefore, is reduced. Consequently, the luminosity is enhanced by about 3%, as measured by the ZDC coincidence rate. A similar observation was made in the He3 + Au run.

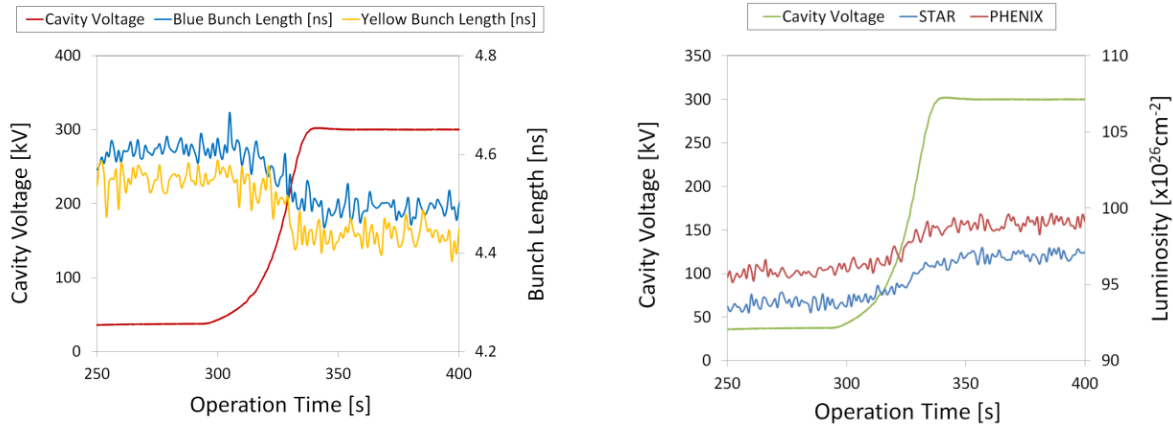

Figure 16: Reduction of the bunch length of both beams in RHIC (left) and the resulting improvement in luminosity (right) observed as the 56 MHz cavity voltage was slowly ramped up.

Beam simulations were performed for the 56 MHz cavity voltages of 300 kV and 2 MV. The simulated luminosity and bunch pattern are shown in Figure 17. Effects included in simulations are intra-beam scattering, cavity's wake-fields, and stochastic cooling. Parameters used in these simulations reflect those during the normal operation of RHIC. The results indicate a 5% luminosity increase when the 300 kV is applied to the beam. This agrees well with our experimental observations. The same simulation tool also predicts that the luminosity will be ~30% higher if the full 2 MV gap voltage is reached.

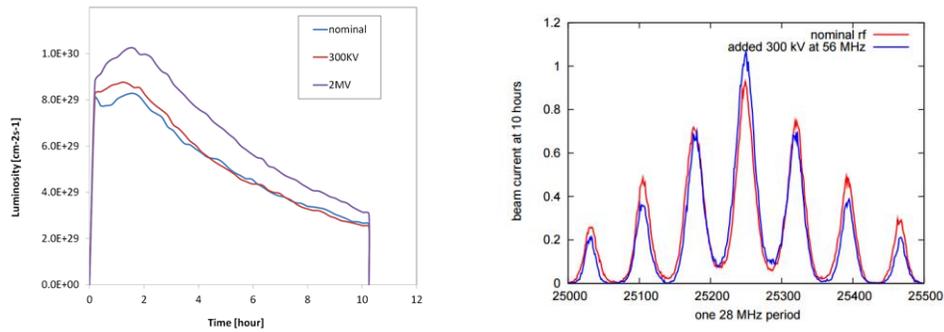

Figure 17: Simulations of the RHIC luminosity (top) and bunch pattern (bottom) at different voltages of the 56 MHz cavity.

The simulations also predict change of the population of ions in different 197 MHz RF buckets when Au beam is re-bucketed. Population of the particles in the satellite buckets is reduced because the bunch length is squeezed by the 56 MHz cavity. We also observed this population difference, between turning on/off the 56 MHz cavity at the end of the store time (about ten hours), using a wall current monitor, as shown in bottom of Figure 18. A decrease in the bunch length is 4%.

With full current beam in RHIC, a total power extracted from the single HOM coupler was measured as 4.1 W. An extracted power at the fundamental frequency was 2.5 W, corresponding to an external $Q$ of $4.1 \times 10^{10}$. All HOMs under 600 MHz with a power of 0.8 mW or higher were identified using a spectrum analyzer; the HOMs at higher frequencies were below the measurement resolution. No measurable degradation of the RHIC beam due to the HOMs from the 56 MHz cavity was observed.

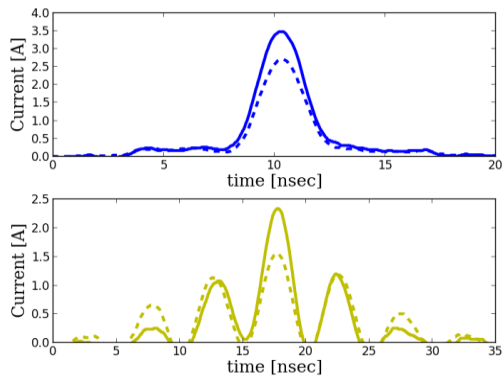

Figure 18: The longitudinal profile of the He3 beam (in blue ring, top) and the Au beam (in yellow ring, bottom). The solid lines and dashed lines correspond to operation with the 56 MHz cavity on and off, respectively.

### C. Helium Processing

Helium processing is a well-developed technique for SRF cavities over decades [13], and successful demonstrations of cavity performance improvement have been approved at various facilities [14, 15].

The large inner volume of the 56 MHz cavity makes it difficult to establish a high quality RF surface after buffered chemical processing and high pressure rinsing. Along with high power conditioning, *in situ* helium processing is also used to achieve high RF gradient. We established a helium partial pressure of $1\times10^{-5}$ Torr in the cavity, while baseline pressure was measured below 2x10-10 Torr at 4 K. Helium gas was introduced into the cavity via a turbo pumping port, and the gas inlet valve was precisely controlled and monitored. The cavity operated with gradually increasing RF power during the addition of the helium gas for 24 hours accumulated. The conditioning alternated between pulsed and continuous wave RF input signals.

The peak cavity gradient level with low field emission improved by 33% from the helium-ion surface cleaning affect alone.

## III.   Conclusion

We successfully designed, installed, and commissioned the first QWR SRF cavity in RHIC. The initial beam tests showed a luminosity boosting of 3%, and a bunch length shrinking of 4% at 1/6 of the design cavity voltage. The HOMs were very efficiently damped under full beam current. The higher order mode coupler will be redesigned to ensure that full voltage can be achieved in RHIC during run 2016, and so experiments can realize a 30% boost in luminosity as expected from theoretical studies.

## Acknowledgement

This works is supported by by Brookhaven Science Associates, LLC under Contract No. DE-AC02-98CH10886 with the U.S. DOE. The authors would like to acknowledge all the support from cavity fabrication, post processing, and most of all, the RHIC control room support on the cavity operation. The authors would like to thank Gary McIntyre, Scott Seberg, Richard Tallman, and Harold Dorr on the processing, assembly, and installation of the 56 MHz cavity cryomodule.